\documentclass[a4paper,11pt]{article}
\usepackage[DIV12]{typearea}
\usepackage{longtable}
\usepackage{booktabs}
\usepackage{amsmath}
\usepackage{amssymb}
\usepackage{bbm}
\usepackage[utf8]{inputenc}
\usepackage{pdflscape}
\usepackage{xspace}
\usepackage[caption=false]{subfig}
\usepackage{nicefrac}
\usepackage{graphicx}
\usepackage{cite}  
\usepackage{nccfoots}
\usepackage{cancel}
\usepackage[small]{caption2}
\usepackage{multirow}
\usepackage{extarrows}
 
\newcommand{\subind}[1]{{{\ensuremath\scriptscriptstyle{(\hspace{-0.7pt}#1\hspace{-0.7pt})}}}}

\newcommand{\rep}[1]{\ensuremath\boldsymbol{#1}}

\newcommand{\x}{\ensuremath\times}
\newcommand{\Z}[1]{\ensuremath{\mathbbm{Z}_{#1}}} 

\newcommand{\SL}[1]{\ensuremath{\mathrm{SL}(#1)}}

\newcommand{\vev}[1]{\ensuremath{\langle{#1}\rangle}}
\newcommand{\I}{\mathrm{i}}
\newcommand{\Id}{\mathbbm{1}}

\newcommand{\CP}{\ensuremath{\mathcal{CP}}\xspace}

\newcommand{\trafo}[1]{\ensuremath{\stackrel{#1}{\longrightarrow}}}

\newcommand{\nphantom}[1]{\sbox0{#1}\hspace{-\the\wd0}}

\usepackage{xcolor}

\advance \headheight by 3.0truept       
\usepackage[pdftex]{hyperref}

\hypersetup{
    pdftitle = {Eclectic flavor scheme from ten-dimensional string theory - I. Basic results},
    pdfauthor = {Nilles, Ramos-Sanchez, Vaudrevange}
}
 
\begin{document}

\begin{titlepage}

\begin{flushright}
\normalsize{TUM-HEP 1266/20}
\end{flushright}

\vspace*{1.0cm}

\begin{center}
{\Large\textbf{\boldmath Eclectic flavor scheme from ten-dimensional string theory - I\\ Basic results\unboldmath}}
\vspace{1cm}

\textbf{%
Hans Peter Nilles$^{a}$, Sa\'ul Ramos--S\'anchez$^{b,c}$, Patrick K.S. Vaudrevange$^{c}$
}
\\[8mm]
\textit{$^a$\small Bethe Center for Theoretical Physics and Physikalisches Institut der Universit\"at Bonn,\\ Nussallee 12, 53115 Bonn, Germany}
\\[2mm]
\textit{$^b$\small Instituto de F\'isica, Universidad Nacional Aut\'onoma de M\'exico,\\ POB 20-364, Cd.Mx. 01000, M\'exico}
\\[2mm]
\textit{$^c$\small Physik Department T75, Technische Universit\"at M\"unchen,\\ James-Franck-Stra\ss e 1, 85748 Garching, Germany}
\end{center}

\vspace*{1cm}

\begin{abstract}
In a consistent top-down approach based on orbifold compactifications, modular and traditional 
flavor symmetries combine nontrivially to the so-called eclectic flavor symmetry. We extend this 
scheme from two extra dimensions, discussed previously, to the six extra dimensions of string 
theory. By doing so, new insights on the nature of $\CP$ and its spontaneous breaking emerge. 
Moreover, we identify a new interpretation of $R$-symmetries as unbroken remnants from modular 
symmetries that are associated with geometrically stabilized complex structure moduli. Hence, all 
symmetries (i.e.\ modular, traditional flavor, $\CP$ and $R$) share a common origin in string 
theory: on a technical level, they are given by outer automorphisms of the Narain space group. The 
eclectic top-down approach leads to a very restrictive scheme with high predictive power. It 
remains a challenge to connect this with existing bottom-up constructions of modular flavor 
symmetry.
\end{abstract}

\end{titlepage}

\newpage

\section{Introduction}

We present a study of the eclectic flavor approach~\cite{Nilles:2020nnc,Nilles:2020kgo} from the 
point of view of ten-dimensional (10d) string theory with six compact dimensions. Eclectic flavor 
groups appear naturally in string constructions from the nontrivial product of modular and 
traditional flavor groups. This work extends previous analyses~\cite{Baur:2019kwi,Baur:2019iai} 
that were based on a simplified picture with two extra dimensions (where the four additional 
compact dimensions of 10d string theory were treated as independent spectators), which 
apparently does not include all aspects of the full 10d top-down discussion. New information is 
especially relevant in view of \CP and, moreover, $R$-symmetries that appear as remnants of the 
Lorentz group in six extra dimensions. With these $R$-symmetries, we discover that the $R$-charges 
are intrinsically related to the modular weights of the corresponding matter fields. The full 
picture leads to an enlargement of the eclectic flavor group as well as its subgroups that 
represent enhanced local flavor symmetries at specific points (or other sub-loci) in moduli space. 
These enhanced symmetries appear due to the alignment of vacuum expectation values (vevs) of 
so-called moduli fields and might include \CP-like symmetries. They are spontaneously broken at 
generic regions in moduli space.

In this paper we present the outcome of our analysis without going into technical details. We shall 
illustrate our results in a specific benchmark model with a sub-sector based on the 
$\mathbbm T^2/\Z3$ orbifold. The generic situation and the full technical discussion will be 
relegated to an upcoming companion paper~\cite{Nilles:2020pr}. In section~\ref{sec:overview} we 
shall review and summarize the results of the eclectic approach that are known so far. These 
include the origin of traditional and modular flavor symmetries, their inclusion in the eclectic 
picture, as well as the restrictions on the low-energy effective action represented by the 
superpotential and the K\"ahler potential. Section~\ref{sec:flavorRSymmetries} will be devoted to 
the completion of flavor symmetries within the full six-dimensional orbifold picture. These are 
mainly due to remnant symmetries of the modular group of the complex structure moduli and they 
enhance the eclectic flavor symmetry by $R$-symmetries. Moreover, the nature of \CP-like 
transformations is only fully uncovered in six (and not two) extra dimensions. In 
section~\ref{sec:LocalFlavorUnification} we shall discuss the enhancements of the flavor groups at 
specific locations in moduli space. Section~\ref{sec:conclusions} will give conclusions and outlook 
for future research.

\vspace{-4mm}
\section{Status of flavor symmetries in string orbifolds}
\label{sec:overview}
\enlargethispage{\baselineskip}

\paragraph{Traditional flavor symmetries.} 
Discrete flavor symmetries appear naturally in string theories~\cite{Kobayashi:2006wq} with the low 
energy particle content of the Standard Model of particle physics~\cite{Olguin-Trejo:2018wpw}. 
Furthermore, they are instrumental as an explanation of mass hierarchies and mixings of quarks and 
leptons in bottom-up extensions of the Standard Model~\cite{Feruglio:2019ktm}. In particular, 
flavor symmetries based on non-Abelian finite groups are promising, especially if they allow for 
three-dimensional irreducible representations corresponding to the three generations of quarks 
and/or leptons. A crucial step in flavor model building is the (spontaneous) breaking of the 
non-Abelian flavor symmetry. Traditionally, this breaking is achieved by the vevs of scalars, 
called flavon fields, that are uncharged under the Standard Model but transform in a 
higher-dimensional representation of the flavor symmetry.

\paragraph{Modular symmetries.} Recently, a new bottom-up approach based on modular groups and 
their realizations as finite modular groups has become popular, inspired by the seminal work of 
Feruglio~\cite{Feruglio:2017spp}. The modular group $\SL{2,\Z{}}$ can be defined by two abstract 
generators $\mathrm{S}$ and $\mathrm{T}$, subject to the relations
\begin{equation}
\label{eq:DefiningRelationsOfSL2Z}
\mathrm{S}^4 ~=~ (\mathrm S\,\mathrm T)^3 ~=~ \Id,\qquad \mathrm S^2\,\mathrm T ~=~ \mathrm T\,\mathrm S^2 \,.
\end{equation}
A choice of its generators $\mathrm{S}$ and $\mathrm{T}$ is given by
\begin{equation}\label{eq:SL2ZGeneratorsSandT}
\mathrm{S} ~:=~ \left(\begin{array}{cc} 0 & 1\\ -1 & 0\end{array}\right) \qquad\text{and}\qquad 
\mathrm{T} ~:=~ \left(\begin{array}{cc} 1 & 1\\ 0 & 1\end{array}\right)\;.
\end{equation}
In these terms, all elements $\gamma\in\SL{2,\Z{}}$ are expressed as $2\times 2$ matrices with 
integer entries and unit determinant. The modular group $\SL{2,\Z{}}$ is accompanied by a special 
scalar field, called the modulus $M$, that transforms as
\begin{equation}\label{eq:ActionOfSL2ZOnT}
M ~\stackrel{\mathrm{S}}{\longrightarrow}~ -\frac{1}{M} \qquad\mathrm{and}\qquad M ~\stackrel{\mathrm{T}}{\longrightarrow}~ M+1\;.
\end{equation}
Moreover, the modular group $\SL{2,\Z{}}$ acts on matter fields with unitary representation 
matrices that generate a finite group, the so-called finite modular group. For example, the finite 
modular groups $\Gamma'_N$ for $N \in\{2,3,4,5\}$ can be defined by the abstract 
definition~\eqref{eq:DefiningRelationsOfSL2Z} combined with the additional relation 
$\mathrm{T}^N = \Id$, see ref.~\cite{Baur:2019kwi} for $N=3$ and ref.~\cite{Liu:2019khw} for 
general $N$. In the scheme of modular symmetries, the crucial step of flavon alignment from the 
traditional approach finds a natural explanation through couplings $\hat{Y}(M)$ that depend solely 
on the modulus $M$ but still transform in higher-dimensional representations of the finite 
modular group: couplings become so-called modular forms of $M$. Thus, the alignment of the coupling 
$\hat{Y}(M)$ in flavor space depends only on the vev $\langle M\rangle$ of the single modulus $M$. 
Interestingly, the modular approach can be naturally extended to include $\CP$-like transformations 
such that the vev $\langle M\rangle$ can also be a measure of spontaneous $\CP$ breaking, see 
refs.~\cite{Dent:2001cc,Baur:2019kwi,Novichkov:2019sqv}.

\paragraph{Unified picture of traditional and modular flavor symmetry in string theory.} The 
approach based on a traditional flavor symmetry and the one utilizing a (finite) modular symmetry 
seem orthogonal from a bottom-up perspective. One can distinguish between two cases: a symmetry 
transformation can either leave the modulus invariant, giving rise to a traditional flavor 
transformation, or transform the modulus nontrivially, yielding a modular transformation.

However, in a unified picture based on string theory, they turn out to be intimately related. First 
of all, it is well-known that string theory compactified on a two-torus $\mathbbm{T}^2$ is equipped 
with two modular groups $\SL{2,\Z{}}_T$ and $\SL{2,\Z{}}_U$ associated with two moduli: the 
so-called K\"ahler modulus $T$ and the complex structure modulus $U$, respectively. Furthermore, in 
toroidal orbifold compactifications, both, traditional and modular symmetries, originate from outer 
automorphisms of the so-called Narain space group~\cite{Baur:2019kwi}, which encodes a string 
compactification on an orbifold geometry, see ref.~\cite{GrootNibbelink:2017usl} for a technical 
discussion of Narain space groups. This avenue of outer automorphisms of the Narain space group 
was initially discussed in a general study of two-dimensional orbifolds 
$\mathbbm{T}^2/\Z{N}$~\cite{Baur:2019kwi,Baur:2019iai}, where the four orthogonal compact 
dimensions are considered as spectators. In what follows we shall illustrate the results in a 
specific example with $N=3$.

\paragraph{\boldmath Symmetries of the $\mathbbm{T}^2/\Z{3}$ orbifold.\unboldmath} 
In this case, the complex structure modulus is stabilized at 
$\vev U=\omega:=\exp\left(\nicefrac{2\pi\I}{3}\right)$, such that the two-torus $\mathbbm{T}^2$ 
exhibits a $\Z{3}$ rotational symmetry. For the $\mathbbm{T}^2/\Z{3}$ orbifold, the modular 
group $\SL{2,\Z{}}_U$ associated with the complex structure modulus $U$ is broken. However, there 
remain several unbroken symmetries, as summarized in table~\ref{tab:Z3FlavorGroups}: there are two 
generators of translational outer automorphisms of the $\mathbbm{T}^2/\Z{3}$ Narain space group, 
denoted by $\mathrm{A}$ and $\mathrm{B}$. Since translations leave all moduli invariant, they 
belong to the traditional flavor symmetry. It turns out that $\mathrm{A}$ and $\mathrm{B}$ generate 
$\Delta(27)$. Then, depending on the geography of strings in extra dimensions, matter fields build 
various representations of $\Delta(27)$: strings that live in the bulk of the extra dimensions are 
singlets, while so-called twisted strings that are localized in extra dimensions at the fixed 
points of the $\mathbbm{T}^2/\Z{3}$ orbifold are triplets or anti-triplets. In addition, 
there exists an unbroken rotational outer automorphism of the $\mathbbm{T}^2/\Z{3}$ Narain space 
group, denoted by $\mathrm{C}$, which leaves the K\"ahler modulus invariant and thus belongs to the 
traditional flavor symmetry. $\mathrm{C}$ enlarges $\Delta(27)$ to $\Delta(54)$, which is known to 
be the flavor symmetry of the $\mathbbm{T}^2/\Z{3}$ orbifold~\cite{Kobayashi:2006wq}. 
However, it is important to note that $\mathrm{C}$ is realized as a $180^\circ$ rotation in the 
extra dimensions of the $\mathbbm{T}^2/\Z{3}$ orbifold~\cite{Baur:2019kwi}, i.e.\ $\mathrm{C}$ is 
an unbroken remnant of the higher-dimensional Lorentz symmetry. Consequently, $\Delta(54)$ is an 
example of a non-Abelian discrete $R$-symmetry of $\mathcal{N}=1$ supersymmetry~\cite{Chen:2013dpa}
arising from string theory, where the superpotential transforms as a nontrivial singlet $\rep{1}'$ 
of $\Delta(54)$~\cite{Baur:2019iai}. 

Similar to the toroidal case, the outer automorphisms of the $\mathbbm{T}^2/\Z{3}$ Narain space 
group include transformations that act nontrivially on the K\"ahler modulus $T$. One can identify 
two generators, denoted also by $\mathrm{S}$ and $\mathrm{T}$, which satisfy the defining 
relations~\eqref{eq:DefiningRelationsOfSL2Z} of the modular group $\SL{2,\Z{}}_T$. When acting on 
localized matter fields, this modular group is realized as the finite modular group 
$T'\cong\Gamma'_3\cong\SL{2,3}$. Therefore, matter fields transform under modular transformations 
in representations of $T'$~\cite{Lauer:1989ax,Lerche:1989cs}. For example, the $\Delta(54)$ triplet 
$\Phi_{\nicefrac{-2}{3}}$ of twisted strings localized at the three fixed points of the 
$\mathbbm{T}^2/\Z{3}$ orbifold transform as $\rep2'\oplus\rep1$ of $T'$ and the subscript 
$\nicefrac{-2}{3}$ of $\Phi_{\nicefrac{-2}{3}}$ gives the $\SL{2,\Z{}}_T$ modular weight (see 
ref.~\cite{Nilles:2020kgo} for details).

\paragraph{The eclectic group and local flavor unification.} The traditional flavor group 
$\Delta(54)$ does not commute with the finite modular group $T'$. This fact gives rise to the 
so-called eclectic flavor group~\cite{Nilles:2020nnc,Nilles:2020kgo}: the nontrivial combination of 
traditional flavor and finite modular groups, where the finite modular group corresponds to outer 
automorphisms of the traditional flavor group. For the $\mathbbm{T}^2/\Z{3}$ orbifold one 
obtains the eclectic flavor group as the multiplicative closure of $\Delta(54)$ and $T'$, i.e.
\begin{equation}
\Omega(1) ~\cong~ [648,533] ~\cong~ \Delta(54) ~\cup~ T'\;,
\end{equation}
where $[648,533]$ denotes the GAP ID~\cite{GAP4} of $\Omega(1)$. Note that the eclectic flavor 
group has a similar structure as the one discussed in ref.~\cite{Ohki:2020bpo}. Some 
phenomenological properties of $\Omega(1)$ have been studied~\cite{Yao:2015dwa,King:2016pgv}, 
considering this group as a traditional flavor symmetry of bottom-up models. In contrast, in our 
eclectic approach, $\Omega(1)$ is nonlinearly realized on the K\"ahler modulus $T$, except for the 
subgroup $\Delta(54)$, as one can see from the action eq.~\eqref{eq:ActionOfSL2ZOnT} on the modulus 
$M=T$. However, there are certain points (or regions, in general) in moduli space, where larger 
subgroups of $\Omega(1)$ are realized linearly: for certain vevs of $T$, some modular 
transformations leave $\langle T\rangle$ invariant and remain unbroken. They give rise to the 
stabilizer subgroup $H_{\langle T\rangle}\subset\SL{2,\Z{}}_T$. Then, embedding the stabilizer 
subgroup into $T'$ yields an enhancement of the traditional flavor symmetry $\Delta(54)$ to a 
so-called unified flavor symmetry at some specific values of $\langle T\rangle$ in moduli space. 
In detail, one can identify two independent points $\langle T\rangle$ with a nontrivial stabilizer 
subgroup $H_{\langle T\rangle}$: 
\begin{subequations}
\label{eqs:H_T}
\begin{eqnarray}
\label{eqs:H_T-T=i}
\mathrm{at}\  \langle T \rangle ~=~ \I     & : & H_{\I}     ~=~ \{\mathrm{S}^k \mathrm{\ for\ } k=0,1,2,3\} \hspace{57,5pt}  ~\cong~ \Z{4}\;, \\
\label{eqs:H_T-T=omega}
\mathrm{at}\  \langle T \rangle ~=~ \omega & : & H_{\omega} ~=~ \{(\mathrm{S}\,\mathrm{T})^k \mathrm{\ for\ } k=0,1,2\} \times \{\Id, \mathrm{S}^2\} ~\cong~ \Z{3}\times\Z{2}\;,
\end{eqnarray}
\end{subequations}
see also refs.~\cite{Novichkov:2018ovf,Gui-JunDing:2019wap}. In both cases, a $\Z{2}$ subgroup is 
generated by $\mathrm{S}^2$. On the level of outer automorphisms of the Narain space group, 
$\mathrm{S}^2$ coincides with the transformation $\mathrm{C}$ discussed earlier and thus belongs to 
the traditional flavor symmetry $\Delta(54)$. Therefore, the enhancement of $\Delta(54)$ by the 
stabilizer subgroup increases the order of the group only by a factor of two (three) in the first 
(second) case, resulting in the following enhancements~\cite{Nilles:2020nnc}:
\begin{subequations}
\label{eqs:localflavorunifiedgroups}
\begin{eqnarray}
\label{eqs:localflavorunifiedgroups-T=i}
\mathrm{at}\  \langle T \rangle ~=~ \I     & : & \Delta(54) ~\rightarrow~ \Sigma(36\times 3) ~\cong~ [108,15]~\subset~ \Omega(1)\;,\\
\label{eqs:localflavorunifiedgroups-T=omega}
\mathrm{at}\  \langle T \rangle ~=~ \omega & : & \Delta(54) ~\rightarrow~ \tilde{Y}(0) \hspace{23,5pt} ~\cong~ [162,10] ~\subset~ \Omega(1)\;,
\end{eqnarray}
\end{subequations}
using ref.~\cite{Jurciukonis:2017mjp} for naming conventions of finite groups. This peculiar 
feature of eclectic flavor models has been named local flavor unification, as different (enhanced) 
flavor symmetries are realized at different points in moduli space. Further enhancements are 
possible if $\CP$ is taken into account as well~\cite{Baur:2019kwi,Baur:2019iai}.

\begin{table}[t!]
\center
\begin{tabular}{|c|c||c|c|c|c|c|}
\hline
\multicolumn{2}{|c||}{nature}                & outer automorphism       & \multicolumn{4}{c|}{\multirow{2}{*}{flavor groups}} \\
\multicolumn{2}{|c||}{of symmetry}           & of Narain space group    & \multicolumn{4}{c|}{}\\
\hline
\hline
\parbox[t]{3mm}{\multirow{5}{*}{\rotatebox[origin=c]{90}{eclectic}}} &\multirow{2}{*}{modular}  & rotation $\mathrm{S}~\in~\SL{2,\Z{}}_T$ & $\Z{4}$      & \multicolumn{2}{c|}{\multirow{2}{*}{$T'$}} &\multirow{5}{*}{$\Omega(1)$}\\
                  &                          & rotation $\mathrm{T}~\in~\SL{2,\Z{}}_T$ & $\Z{3}$      & \multicolumn{2}{c|}{}                      & \\
\cline{2-6}
       & \multirow{3}{*}{traditional flavor} & translation $\mathrm{A}$                & $\Z{3}$      & \multirow{2}{*}{$\Delta(27)$} & \multirow{3}{*}{$\Delta(54)$} & \\
       &                                     & translation $\mathrm{B}$                & $\Z{3}$      &                               & & \\
\cline{3-5}
       &                                     & rotation $\mathrm{C}=\mathrm{S}^2~\in~\SL{2,\Z{}}_T$ & \multicolumn{2}{c|}{$\Z2^R$}  & & \\
\hline
\end{tabular}
\caption{Eclectic flavor group $\Omega(1)\cong\Delta(54)\cup T'\cong [648, 533]$ for a 
$\mathbbm T^2/\Z{3}$ orbifold. Note that the order-3 translational outer automorphisms $\mathrm{A}$ 
and $\mathrm{B}$ generate $\Delta(27)$, which contains the $\Z{3}\times\Z{3}$ point and space group 
selection rules~\cite{Hamidi:1986vh,Dixon:1986qv,Ramos-Sanchez:2018edc}. The rotational outer 
automorphism $\mathrm{C}$ is special as it belongs to both, the traditional flavor symmetry 
$\Delta(54)$ and the finite modular symmetry $T'$.}
\label{tab:Z3FlavorGroups}
\end{table}

\paragraph{Eclectic restrictions on superpotential and K\"ahler potential.} 
The eclectic flavor group severely constrains the allowed couplings in the theory for both the 
superpotential and the K\"ahler potential~\cite{Nilles:2020kgo}. Let us illustrate 
this in a specific example. If one demands a modular symmetry $\SL{2,\Z{}}_T$ with finite modular 
group $T'$, the superpotential has to transform as
\begin{equation}
\label{eq:modularSL2ZTTrafoOfW}
 \mathcal W ~\stackrel{\gamma_{\scriptscriptstyle T}}{\longrightarrow}~ (c\,T+d)^{-1}\,\mathcal W\;,\qquad\mathrm{under}\quad
 \gamma_{\scriptscriptstyle T}~=~\left(\begin{array}{cc}a&b\\c&d\end{array} \right) ~\in~\SL{2,\Z{}}_T\,.
\end{equation}
In other words, even though the superpotential $\mathcal W$ does transform nontrivially under 
$\gamma_{\scriptscriptstyle T}\in\SL{2,\Z{}}_T$, it is invariant under $T'$, where matter fields 
and Yukawa couplings $\hat{Y}(T)$ transform with automorphy factors $(c\,T+d)^n$ with modular 
weights $n$ under $\SL{2,\Z{}}_T$ and, simultaneously, in representations of $T'$. Now, we consider 
the trilinear superpotential $\mathcal W$ of three copies of twisted strings $\Phi_{\nicefrac{-2}{3}}^i$, 
where $n=\nicefrac{-2}{3}$ and $i=1,2,3$, localized at the three fixed points of the $\mathbbm T^2/\Z{3}$ 
orbifold~\cite{Nilles:2020kgo}. If we impose only invariance under the finite modular 
group $T'$ the superpotential $\mathcal W$ can be parameterized by four coefficients. These 
correspond to four independent couplings that are undetermined from a bottom-up perspective 
based solely on the modular flavor symmetry. If we now consider the full eclectic group $\Omega(1)$, 
which includes the extension of $T'$ by the traditional flavor symmetry $\Delta(54)$, these four 
coefficients from the $T'$ theory are forced to be identical and there remains only one independent 
parameter. Similarly, the bilinear K\"ahler potential (which is essentially unconstrained in 
bottom-up models based solely on (finite) modular symmetries~\cite{Chen:2019ewa}) is very sensitive 
to the eclectic extension: starting with only $T'$, the K\"ahler potential of localized strings 
contains non-diagonal terms, which are removed by the traditional flavor symmetry $\Delta(54)$ to 
all orders in the K\"ahler modulus $T$. Therefore, eclectic flavor symmetries are highly predictive and 
their phenomenological implications are worth to be studied in more detail.

\section{\boldmath New insights from six compact dimensions\unboldmath}
\label{sec:6DInsights}

The eclectic approach to flavor symmetries has been discovered by the study of two-dimensional 
$\mathbbm T^2/\Z{N}$ orbifolds and discussed in detail for $N=3$. As we show next, the 
generalization to six extra dimensions might seem straightforward, but it still yields new insights 
concerning $R$-symmetries and \CP.

\subsection{\boldmath Flavor $R$-symmetries from modular symmetries\unboldmath}
\label{sec:flavorRSymmetries}

It is known that compactifications of heterotic strings on factorizable orbifolds $\mathbbm{T}^6/P$ 
lead to various discrete $R$-symmetries~\cite{Kobayashi:2004ya,Bizet:2013gf,Nilles:2013lda,Bizet:2013wha,Nilles:2017heg}. 
They can be understood as discrete remnants of the originally ten-dimensional Lorentz symmetry that 
are described by so-called {\it sublattice rotations} in the compactified extra dimensions. A 
sublattice rotation $\mathrm{R}$ is defined as a discrete rotational symmetry of the orbifold that 
is not part of the point group, i.e.\ $\mathrm{R}\not\in P$. As such, $\mathrm{R}$ has to map the 
torus $\mathbbm{T}^6$ underlying the orbifold $\mathbbm{T}^6/P$ to itself. For factorizable 
orbifolds, the six-torus is a direct product $\mathbbm{T}^6 = \mathbbm{T}^2\times\mathbbm{T}^2\times\mathbbm{T}^2$, 
where an Abelian point group $P$ acts diagonally in the three two-tori. Then, a separate rotation 
only in one $\mathbbm T^2$ sub-torus of $\mathbbm{T}^6$ is a symmetry of the orbifold geometry. For 
instance, the factorizable $\mathbbm{T}^6/\Z6$-II orbifold geometry allows three sublattice 
rotations: a separate $\Z6$, $\Z3$ and $\Z2$ rotation in the first, second and third two-torus, 
respectively. It turns out that under these discrete rotations localized strings carry fractional 
charges. Taking these fractional charges into account, the three sublattice rotations yield a 
$\Z{36}^R\x\Z{9}^R\x\Z{4}^R$ $R$-symmetry for the $\mathbbm{T}^6/\Z6$-II orbifold 
geometry~\cite{Nilles:2013lda} (where we give the order of the $R$-symmetries for the superfields; 
in this normalization, fermions can have half-integer $R$-charges). In this section, we 
present an alternative interpretation of these discrete $R$-symmetries using the example of the 
$\Z{9}^R$ $R$-symmetry associated with the $\mathbbm T^2/\Z3$ orbifold sector (corresponding to the 
second torus of the $\mathbbm{T}^6/\Z6$-II orbifold) discussed above: we show that sublattice 
rotations and, hence, their corresponding discrete $R$-symmetries can be understood as discrete 
remnants of modular symmetries. In this way, $R$-symmetries appear on the same footing as modular 
and flavor symmetries, with common origin in the outer automorphisms of the Narain space group. 
Moreover, it is obvious that these $R$-symmetries are linked to the Lorentz group of the six-torus, 
and we would like to understand why these symmetries had not been identified in the earlier 
discussions~\cite{Baur:2019kwi,Baur:2019iai} that just considered an (orbifolded) two-torus. In the 
following we shall clarify this situation. 

A two-torus $\mathbbm T^2$ can be parameterized by a K\"ahler modulus $T$ and a complex structure 
modulus $U$. For the $U$ modulus there are two outer automorphisms of the corresponding Narain 
lattice, denoted here by $\mathrm S_{\scriptscriptstyle U}$ and $\mathrm T_{\scriptscriptstyle U}$, 
which satisfy the defining relations~\eqref{eq:DefiningRelationsOfSL2Z} of $\SL{2,\Z{}}_U$. They 
act on the modulus as $U\trafo{\mathrm S_{\scriptscriptstyle U}}-1/U$ and 
$U\trafo{\mathrm T_{\scriptscriptstyle U}}U+1$, while leaving $T$ untouched. One can show that this 
action translates into the transformations
\begin{equation}
\label{eq:SL2ZOnT2Basis}
e_1  ~\stackrel{\mathrm S_{\scriptscriptstyle U}}{\longrightarrow}~ -e_2\,,\quad e_2 ~\stackrel{\mathrm S_{\scriptscriptstyle U}}{\longrightarrow}~ e_1\,,\qquad\mathrm{and}\qquad
e_1  ~\stackrel{\mathrm T_{\scriptscriptstyle U}}{\longrightarrow}~  e_1\,,\quad e_2 ~\stackrel{\mathrm T_{\scriptscriptstyle U}}{\longrightarrow}~ e_1+e_2
\end{equation}
of the $\mathbbm T^2$ basis vectors $e_1$ and $e_2$. In addition, similar to 
eq.~\eqref{eq:modularSL2ZTTrafoOfW}, the superpotential transforms as
\begin{equation}
\label{eq:modularSL2ZUTrafoOfW}
 \mathcal W ~\stackrel{\gamma_{\scriptscriptstyle U}}{\longrightarrow}~ (c\,U+d)^{-1}\,\mathcal W\,,\qquad\text{under}\quad
  \gamma_{\scriptscriptstyle U}~=~\left(\begin{array}{cc}a&b\\c&d\end{array} \right) \in\SL{2,\Z{}}_U\,.
\end{equation}

Next, we observe from eq.~\eqref{eq:SL2ZOnT2Basis} that we can express a $\Z{N}$ rotational 
symmetry $\mathrm{R}$ of the two-torus as a modular transformation 
$\mathrm{R}=\gamma_\subind{N}\in\SL{2,\Z{}}_U$ of order $N$, i.e.\ $(\gamma_\subind{N})^N=\Id$. 
This modular transformation $\gamma_\subind{N}$ acts on the superpotential as in 
eq.~\eqref{eq:modularSL2ZUTrafoOfW}. However, the automorphy factor of a $\Z{N}$ rotation 
$\gamma_\subind{N}$ is just a phase of order $N$ because $(\gamma_\subind{N})^N=\Id$ implies 
$(c\,U+d)^{-N}=1$. Here, one already realizes that (for $N\neq 2$) the modulus $U$ has to be 
stabilized to a fixed value \vev{U} such that $(c\,\vev{U}+d)^{-N}=1$ is valid. Indeed, one can 
show that $U$ has to be invariant under a modular transformation that corresponds to a $\Z{N}$ 
rotational symmetry of the two-torus if $N\neq 2$. For example, the modular transformation 
$S_{\scriptscriptstyle U}\,T_{\scriptscriptstyle U}\in\SL{2,\Z{}}_U$ generates a $\Z3$ 
rotation if $e_1$ and $e_2$ have equal lengths and enclose an angle of $120^\circ$, i.e.\ if $U$ is 
stabilized at $\omega=\exp\left(\nicefrac{2\pi\I}{3}\right)$.

Let us now consider the case of a (six-dimensional) orbifold geometry $\mathbbm{T}^6/P$ and focus 
on a sublattice rotation of a $\mathbbm T^2/\Z{N}$ orbifold sector, where $N\in\{2,3,4,6\}$. Due 
to the point group action, the complex structure modulus $U$ of the $\mathbbm T^2/\Z{N}$ orbifold 
sector has to be stabilized geometrically at some value \vev{U} depending on $N\neq 2$ (see 
ref.~\cite{Kikuchi:2020frp} for a related discussion). This vev breaks $\SL{2,\Z{}}_U$ to the 
stabilizer group $H_{\vev U}\subset\SL{2,\Z{}}_U$ that leaves \vev{U} invariant, and the stabilizer 
group $H_{\vev U}$ contains the sublattice rotation $\mathrm{R}=\gamma_\subind{N}$ that maps the 
$\mathbbm T^2/\Z{N}$ orbifold sector to itself. Moreover, due to the nontrivial transformation of 
the superpotential eq.~\eqref{eq:modularSL2ZUTrafoOfW}, each unbroken modular transformation 
$\mathrm{R}=\gamma_\subind{N}$ generates a discrete Abelian $R$-symmetry. Thus, the sublattice 
rotation $\gamma_\subind{N}\in H_{\vev U}\subset\SL{2,\Z{}}_U$ can be identified with an 
automorphism of the Narain space group. Is this transformation an independent symmetry of the 
orbifold? If the orbifold is two-dimensional, $\mathrm R$ belongs to the point group $P$. Hence, 
$\mathrm R$ is an inner automorphism in the two-dimensional case and, consequently, not an 
independent symmetry of the theory. This is different in the six-dimensional case, as we shall 
explain now. The key observation is that in a six-dimensional orbifold compactification which 
preserves $\mathcal{N}=1$ supersymmetry, the point group consists of rotations that act 
simultaneously on more than two compact dimensions and not only on one $\mathbbm T^2/\Z{N}$ 
orbifold sector. Thus, a two-dimensional sublattice rotation 
$\mathrm{R}=\gamma_\subind{N}\in H_{\vev U}\subset\SL{2,\Z{}}_U$ cannot belong to the point group 
$P$ of a six-dimensional orbifold geometry. Hence, $\mathrm{R}$ is an independent outer 
automorphism of the Narain space group of six-dimensional orbifolds and therefore an authentic 
symmetry of the theory. Since it leaves all K\"ahler and complex structure moduli invariant, it 
contributes to the traditional flavor symmetry.

\begin{table}[t!]
\center
\begin{tabular}{|c|c||c|c|c|c|c|c|}
\hline
\multicolumn{2}{|c||}{nature}        & outer automorphism       & \multicolumn{5}{c|}{\multirow{2}{*}{flavor groups}} \\
\multicolumn{2}{|c||}{of symmetry}   & of Narain space group    & \multicolumn{5}{c|}{}\\
\hline
\hline
\parbox[t]{3mm}{\multirow{6}{*}{\rotatebox[origin=c]{90}{eclectic}}} &\multirow{2}{*}{modular}            & rotation $\mathrm{S}~\in~\SL{2,\Z{}}_T$ & $\Z{4}$      & \multicolumn{3}{c|}{\multirow{2}{*}{$T'$}} &\multirow{6}{*}{$\Omega(2)$}\\
&                                    & rotation $\mathrm{T}~\in~\SL{2,\Z{}}_T$ & $\Z{3}$      & \multicolumn{3}{c|}{}                      & \\
\cline{2-7}
&                                    & translation $\mathrm{A}$                & $\Z{3}$      & \multirow{2}{*}{$\Delta(27)$} & \multirow{3}{*}{$\Delta(54)$} & \multirow{4}{*}{$\Delta'(54,2,1)$} & \\
& traditional                        & translation $\mathrm{B}$                & $\Z{3}$      &                               & & & \\
\cline{3-5}
& flavor                             & rotation $\mathrm{C}=\mathrm{S}^2\in\SL{2,\Z{}}_T$      & \multicolumn{2}{c|}{$\Z{2}^R$} & & & \\
\cline{3-6}
&                                    & rotation $\mathrm{R}=\gamma_\subind{3}\in\SL{2,\Z{}}_U$ & \multicolumn{3}{c|}{$\Z{9}^R$}   & & \\
\hline
\end{tabular}
\caption{Eclectic flavor group $\Omega(2)$ for orbifolds $\mathbbm T^6/P$ that contain a 
$\mathbbm T^2/\Z{3}$ orbifold sector. In this case, $\SL{2,\Z{}}_U$ of the stabilized complex 
structure modulus $U=\exp\left(\nicefrac{2\pi\I}{3}\right)$ is broken, resulting in a remnant 
$\Z{9}^R$ $R$-symmetry. Including $\Z{9}^R$ enhances the traditional flavor group $\Delta(54)$ to 
$\Delta'(54,2,1)\cong [162,44]$ and, thereby, the eclectic group to 
$\Omega(2) \cong [1944, 3448]$. Note that $\Omega(1)\subset \Omega(2)$.}
\label{tab:Z3FlavorGroupsExtended}
\end{table}

Let us now consider our example of a $\mathbbm T^2/\Z3$ orbifold sector embedded in a 
six-dimensional orbifold $\mathbbm T^6/P$. We sketch the main results here and do not enter a 
detailed technical discussion which will be included in ref.~\cite{Nilles:2020pr}. In this case, 
the associated complex structure modulus is fixed, for example, at $\vev{U}=\omega$. Consequently, 
the stabilizer group is $H_{\vev{U}}\cong\Z3\x\Z2$, where the $\Z2$ factor is generated by 
$\mathrm S_{\scriptscriptstyle U}^2$. It is known from the Narain formalism that 
$\mathrm S_{\scriptscriptstyle U}^2 = \mathrm S^2 = \mathrm C$, i.e.\ $\SL{2,\Z{}}_T$ and 
$\SL{2,\Z{}}_U$ share a common element~\cite{Baur:2019iai}. This \Z2 factor is already included in 
both, the traditional flavor symmetry $\Delta(54)$ and the modular symmetry $\SL{2,\Z{}}_T$. On the 
other hand, the \Z3 factor of $H_{\vev U}$ is generated by the $\SL{2,\Z{}}_U$ element 
$\mathrm R:=\mathrm{S}_{\scriptscriptstyle U}\mathrm{T}_{\scriptscriptstyle U}$. It is represented 
by the $2\x2$ matrix 
\begin{equation}
\label{eq:Rgenerator}
\mathrm R~=~\mathrm{S}_{\scriptscriptstyle U}\,\mathrm{T}_{\scriptscriptstyle U}
         ~=~\left(\begin{array}{cc}0&1\\-1&0\end{array}\right) \left(\begin{array}{cc}1&1\\0&1\end{array}\right)
         ~=~\left(\begin{array}{cc}0&1\\-1&-1\end{array}\right) ~\in~ H_{\vev U} ~\subset~ \SL{2,\Z{}}_U\,,
\end{equation}
which acts, according to eq.~\eqref{eq:SL2ZOnT2Basis}, on the torus lattice as a \Z3 sublattice 
rotation: $e_1\trafo{\mathrm R}-e_1-e_2$ and $e_2\trafo{\mathrm R} e_1$. 
Eq.~\eqref{eq:modularSL2ZUTrafoOfW} evaluated at $\vev{U}=\omega$ implies that the 
transformation $\mathrm R$ given in eq.~\eqref{eq:Rgenerator} acts on the superpotential as 
$\mathcal W\trafo{\mathrm R} (-\omega-1)^{-1}\mathcal W = \omega\,\mathcal W$. Modular 
transformations from the stabilizer group $H_{\vev U}$, which include in this case
the \Z3 sublattice rotation 
$\mathrm R$, leave the moduli and hence the K\"ahler potential invariant. Thus, the 
Grassmann number $\vartheta$ must transform under the $\Z3$ sublattice rotation as 
$\vartheta\trafo{\mathrm R}\omega^{1/2}\,\vartheta$, which implies that $\mathrm R$ 
generates a discrete $R$-symmetry. In fact, this discrete $R$-symmetry corresponds to the 
$\Z{9}^R$ symmetry previously identified in ref.~\cite{Nilles:2013lda}. To see this, recall that 
the modular weights $n_{\scriptscriptstyle U}$ of matter fields are multiples of 
$\nicefrac13$, $n_{\scriptscriptstyle U}=\nicefrac{m}{3}$ for $m\in\Z{}$, see e.g.\
ref.~\cite{Ibanez:1992hc,Olguin-Trejo:2017zav}. Consequently, their transformation under 
$\mathrm R$ induces a phase 
\begin{equation}
(c\,\vev U+d)^{n_{\scriptscriptstyle U}}~=~\omega^{\nicefrac{-m}{3}}\,,
\end{equation}
originating from the automorphy factor of $\SL{2,\Z{}}_U$. This means that i) the \Z3 sublattice 
rotational symmetry in the $\mathbbm T^2/\Z3$ orbifold sector of the six-dimensional 
$\mathbbm{T}^6/P$ orbifold is realized as a $\Z{9}^R$ $R$-symmetry, and ii) the discrete 
$R$-charges of matter fields are related to their $\SL{2,\Z{}}_U$ modular weights 
$n_{\scriptscriptstyle U}$. Finally, note that the $\Z{3}$ subgroup of $\Z{9}^R$ is not an 
$R$-symmetry as it maps $\mathcal{W}\trafo{\mathrm R^3}\mathcal{W}$. Moreover, $\mathrm R^3$ 
corresponds to the point group selection rule~\cite{Hamidi:1986vh,Dixon:1986qv,Ramos-Sanchez:2018edc} 
contained already in $\Delta(54)$, see also ref.~\cite{Buchmuller:2008uq}.

These observations lead to an extension of the eclectic flavor group, summarized in 
table~\ref{tab:Z3FlavorGroupsExtended}. On the one hand, the traditional flavor symmetry becomes 
now the multiplicative closure of $\Delta(54)$ and $\Z{9}^R$, which is known as
\begin{equation}
\Delta'(54,2,1) ~\cong~ [162,44] ~\cong~ \Delta(54) ~\cup~ \Z{9}^R\;.
\end{equation}
On the other hand, combining this group with the $T'$ finite modular flavor symmetry arising from 
$\SL{2,\Z{}}_T$, one arrives at the full eclectic flavor symmetry of a $\mathbbm T^2/\Z3$ 
orbifold sector without \CP, being
\begin{equation}
\Omega(2) ~\cong~ \Omega(1) ~\cup~ \Z{9}^R ~\cong~ \Delta'(54,2,1) ~\cup~ T'\;.
\end{equation}
$\Omega(2) \cong [1944,3448]$ can also be written as $\Omega(2)\cong Z(3,2)\rtimes T'$, where 
$Z(3,2)\cong[81,14]$ is generated by $\mathrm A$, $\mathrm B$ and $\mathrm R$.

\subsection{\boldmath \CP for six-dimensional orbifolds $\mathbbm{T}^6/P$\unboldmath}
\label{sec:6DCP}

As shown in refs.~\cite{Baur:2019kwi,Baur:2019iai} for two compact dimensions, the string 
construction of the $\mathbbm T^2/\Z3$ orbifold automatically yields a \Z2 \CP-like transformation 
that maps string states to their \CP-conjugated partner states. In more detail, let us consider a 
string state that is localized in the first twisted sector of the $\mathbbm T^2/\Z3$ orbifold 
geometry. Then, the \CP-partner originates from the second twisted sector. Hence, a \CP-like 
transformation has to be given by an outer automorphism of the Narain space group that interchanges 
the twisted sectors of the $\mathbbm T^2/\Z3$ orbifold. 

How does this situation change for six-dimensional orbifolds $\mathbbm{T}^6/P$? For simplicity, let 
us consider an Abelian point group $P$ that yields exactly three bulk K\"ahler moduli $T_i$, 
$i=1,2,3$, such as $P=\Z{3}\times\Z{3}$, see for example appendix C.1 in ref.~\cite{Fischer:2012qj}. 
These K\"ahler moduli parameterize the sizes (and $B$-fields) of the three two-dimensional planes 
underlying the six-torus $\mathbbm{T}^6$ in which $P$ acts diagonally. Then, each K\"ahler modulus 
$T_i$ is associated with its own modular group $\SL{2,\Z{}}_{T_i}$ for $i=1,2,3$. Moreover, as in 
the two-dimensional discussion, a \CP-like transformation has to interchange simultaneously all 
twisted sectors of the six-dimensional $\mathbbm{T}^6/P$ orbifold that contain \CP-conjugated 
partner states. Consequently, the \CP-like transformation has to act in all six extra dimensions 
simultaneously and not only in two. In this case, one can show that under the \CP-like 
transformation the three K\"ahler moduli transform simultaneously,
\begin{equation}\label{eq:CPinSixDimensions}
T_i  ~\stackrel{\CP}{\longrightarrow}~ -\bar{T}_i\;, \qquad\mathrm{for}\qquad i=1,2,3\;,
\end{equation}
i.e.\ there is no \CP-like transformation that acts nontrivially only on one modulus (e.g.\ 
$T_1 \rightarrow -\bar{T}_1$) while leaving the other moduli invariant (e.g.\ $T_2 \rightarrow T_2$ and 
$T_3 \rightarrow T_3$). Only the combined action eq.~\eqref{eq:CPinSixDimensions} is possible in these 
orbifold constructions (as one might have expected from the holomorphicity of the superpotential). 
An interesting phenomenological consequence is that, if any modulus is stabilized away from a 
so-called self-dual point or region in moduli space, \CP is broken spontaneously.

\section{\boldmath Local flavor unification with modular $R$-symmetries\unboldmath}
\label{sec:LocalFlavorUnification}

As discussed in section~\ref{sec:overview}, at special points $\vev{T}$ in moduli space of 
orbifolds with a $\mathbbm T^2/\Z3$ orbifold sector there is a non-trivial stabilizer group 
$H_{\vev T}\subset\SL{2,\Z{}}_T$ that leaves $\vev{T}$ (as well as the stabilized complex 
structure modulus $\vev U$) invariant. Thus, $H_{\vev T}$ enhances the traditional flavor group to 
a discrete unified flavor group, which is realized linearly as a subgroup of the eclectic flavor 
group. This holds also for the traditional flavor group $\Delta'(54,2,1)\cong [162,44]$ 
which includes $\Delta(54)$ and the remnant $\Z9^R$ $R$-symmetry from $\SL{2,\Z{}}_U$ in the $\mathbbm{T}^2/\Z3$ 
orbifold sector. In this case, one finds different subgroups of the full eclectic flavor group 
$\Omega(2)\cong [1944,3448]$ to be realized linearly for different specific values $\vev T$ of the K\"ahler 
modulus.  Let us consider this enhancement for the two points with non-trivial stabilizer groups discussed in 
eq.~\eqref{eqs:H_T} (in the case where  \CP-like transformations were not taken into account).

According to eq.~\eqref{eqs:H_T-T=i}, the stabilizer group at the point $\langle T\rangle=\I$ is
given by $H_{\vev{T}=\I}\cong\Z4$, generated by the element $\mathrm S$ of $\SL{2,\Z{}}_T$. In this 
case, the traditional flavor group $\Delta'(54,2,1)$ gets enhanced to the unified flavor group 
$\Xi(2,2)\cong[324,111]\subset\Omega(2)$. It is generated by the $\Delta'(54,2,1)$ generators 
$\mathrm A$, $\mathrm B$ and $\mathrm R$, as well as  the stabilizer generator $\mathrm S$ at the 
point $\vev T = \I$. Note that $\mathrm C$ is not an independent generator because it satisfies 
$\mathrm C = \mathrm S^2$. This relation also explains why the order of the unified flavor group is 
just twice (and not four times) the order of the traditional flavor symmetry group. The group 
$\Xi(2,2)$ contains the group $\Sigma(36 \times 3) \cong [108,15]\subset\Omega(1)$ found in the 
absence of the $\Z9^R$ $R$-symmetry (see eq.~\eqref{eqs:localflavorunifiedgroups-T=i}).

At the point $\vev T = \omega$ in moduli space, the unbroken stabilizer group is 
$H_{\vev{T}=\omega} \cong \Z3\x\Z2$ (see eq.~\eqref{eqs:H_T-T=omega}), where only the \Z3 factor, 
generated by $\mathrm{ST}$, is independent since again the \Z2 factor is already included in the 
traditional flavor group as it coincides with $\mathrm C$. Hence, considering the generators 
$\mathrm A$, $\mathrm B$, $\mathrm C$, $\mathrm R$ and $\mathrm{ST}$, the unified flavor group at 
$\vev T=\omega$ is given by $H(3,2,1)\cong[486,125]\subset\Omega(2)$. As before, we find that 
this unified flavor group contains $\tilde Y(0)\cong[162,10]\subset\Omega(1)$ that appears at 
$\vev T=\omega$ when ignoring the existence of the $\Z9^R$ modular remnant symmetry (see 
eq.~\eqref{eqs:localflavorunifiedgroups-T=omega}).

In summary, we find that the traditional flavor group $\Delta'(54,2,1)$ is enhanced at 
different points of moduli space to the following unified flavor groups:
\begin{subequations}
\label{eqs:localflavorunifiedgroups2}
\begin{eqnarray}
\label{eqs:localflavorunifiedgroups2-T=i}
\mathrm{at}\  \langle T \rangle ~=~ \I     & : & \Delta'(54,2,1) ~\rightarrow~ \Xi(2,2) \hspace{16pt}\cong~[324,111]~\subset~ \Omega(2)\;,\\
\label{eqs:localflavorunifiedgroups2-T=omega}
\mathrm{at}\  \langle T \rangle ~=~ \omega & : & \Delta'(54,2,1) ~\rightarrow~ H(3,2,1)~\cong~ [486,125] ~\subset~ \Omega(2)\;,
\end{eqnarray}
\end{subequations}
whose orders, comparing with eqs.~\eqref{eqs:localflavorunifiedgroups}, are enlarged by a factor of 
three because the discrete $\Z9^R$ $R$-symmetry is included. Some of the phenomenological properties
of these unified flavor groups have been addressed in ref.~\cite{King:2016pgv}.

As mentioned earlier, \CP-like transformations are included naturally in string-derived models. 
After stabilizing all moduli, a \Z2 \CP-like transformation remains unbroken if it leaves all 
moduli invariant, cf.\ eq.~\eqref{eq:CPinSixDimensions}. In particular, at the points $\vev T = \I$ 
and $\vev T =\omega$ in K\"ahler moduli space, there exist \CP-like transformations leaving 
$\vev T$ invariant. Thus, assuming that all other moduli are also stabilized at some \CP invariant 
points in moduli space, the unified flavor groups of eq.~\eqref{eqs:localflavorunifiedgroups2} get 
further enhanced. Moreover, there are additional points 
(and lines) beside $\vev T = \I$ and $\vev T =\omega$ that are left invariant by elements of the 
\CP-enhanced modular group, as shown in earlier papers~\cite{Baur:2019kwi,Baur:2019iai}, enriching 
the structure of the possible flavor groups and the possibilities of spontaneous \CP breakdown. We 
shall discuss the details of these possibilities in a companion paper~\cite{Nilles:2020pr}.

\section{Conclusions and outlook}
\label{sec:conclusions}

We have discussed the top-down (TD) approach of eclectic flavor symmetries from the point of view 
of ten-dimensional string theory with six compact spatial dimensions. Compared to previous studies 
that were essentially based on two compact extra dimensions, we can report several new observations:
\begin{itemize}
\item There is a further enlargement of the eclectic flavor group, as illustrated in the difference between 
      table~\ref{tab:Z3FlavorGroups} and table~\ref{tab:Z3FlavorGroupsExtended}.
\item We find a new interpretation of $R$-symmetries connected to both, the modular symmetry of the 
      complex structure modulus and the Lorentz symmetry in six extra dimensions, as explained
      in section~\ref{sec:flavorRSymmetries}.
\item New insights on the nature of \CP-symmetry and its breakdown are discussed in section~\ref{sec:6DCP}.
\item $R$-charges of quark and lepton fields are inherently associated with their $\SL{2,\Z{}}_U$ modular weights.
\end{itemize}
This leads to even more restrictions for TD model building that include strict constraints on the 
K\"ahler potential and the superpotential as well as the possible modular weights and the representations 
of the finite modular flavor group of matter fields. It implies that TD model building is very restrictive, 
with enhanced predictive power. Therefore, it remains a challenge to bridge the gap between the TD approach 
and existing bottom-up constructions of modular flavor symmetries.

\section*{Acknowledgments}

The work of S.R.-S.\ was partly supported by DGAPA-PAPIIT grant IN100217, CONACyT grants F-252167 and 
278017, the Deutsche Forschungsgemeinschaft (SFB1258) and the TUM August--Wilhelm Scheer Program. The 
work of P.V. is supported by the Deutsche Forschungsgemeinschaft (SFB1258).

\providecommand{\bysame}{\leavevmode\hbox to3em{\hrulefill}\thinspace}

\end{document}